\documentclass[preprint]{raa} 
\newcommand{\orcid}[1]{%
    \raisebox{0.7ex}{\scalebox{1}{
        \href{https://orcid.org/#1}{\includegraphics[height=1.5ex]{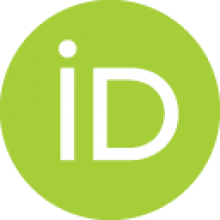}}%
    }}%
}
\usepackage{graphicx,times}             
\usepackage{natbib}
\usepackage{amssymb,amsmath}
\usepackage{array}
\bibpunct{(}{)}{;}{a}{}{,}
\usepackage[pagebackref=true]{hyperref}
\usepackage{multirow}

\usepackage{xcolor}
\usepackage{tcolorbox} 
\tcbuselibrary{breakable} 
\usepackage{soul}

\newif\ifshowrevision
\showrevisiontrue  

\ifshowrevision
  
  \newcommand{\revdel}[1]{\textcolor{red}{\sout{#1}}}
  
\else
  \newcommand{\revdel}[1]{}     
\fi

\newif\ifshowcomments
\showcommentstrue 

\ifshowcomments
  \newcommand{\highlightcomment}[2][yellow!20]{%
    \begin{tcolorbox}[
      colback=#1,            
      colframe=#1,           
      boxrule=0pt,           
      arc=0pt,               
      left=2pt,right=2pt,    
      top=1pt,bottom=1pt,    
      breakable              
    ]
    #2
    \end{tcolorbox}%
  }
\else
  \newcommand{\highlightcomment}[2][yellow!20]{} 
\fi

\begin{document}

  \title{SVOM/VT: On-ground processing of VT--VHF data}

   \volnopage{Vol.0 (202x) No.0, 000--000}      
   \setcounter{page}{1}          

\author{
Chao Wu\orcid{0009-0001-7024-3863}
\inst{1,2}
\and Jesse T. Palmerio\orcid{0000-0002-9408-1563}
\inst{3}
\and Tatyana  Sadibekova\orcid{0000-0002-5162-4222}
\inst{4}
\and Yannis  Canton\orcid{0009-0002-7205-3788}
\inst{5}
\and Kamshat  Tazhenova
\inst{3}
\and  Susanna Diana Vergani\orcid{0000-0001-9398-4907}
\inst{5}
\and Li-Ping Xin\orcid{0000-0002-9422-3437}
\inst{1}
\and Yu-Lei Qiu\orcid{0009-0007-7207-4884}
\inst{1}
\and  Henri Louvin\orcid{0009-0006-1908-4592}
\inst{3}
\and Mo Zhang
\inst{1}
\and Mao-Hai Huang
\inst{1,2}
\and Isabelle  Jegouzo
\inst{6}
\and Hua-Li Li
\inst{1}
\and Hong-bo Cai
\inst{1}
\and Jin-Song  Deng 
\inst{1,2}
\and Bertrand Cordier
\inst{3}
\and Jian-Yan Wei
\inst{1,2}
}
\institute{
National Astronomical Observatories, Chinese Academy of Sciences, Beijing 100101,  China; {\it{cwu@nao.cas.cn}} \\ 
\and 
School of Astronomy and Space Science, University of Chinese Academy of Sciences, Beijing 101408, China \\ 
\and 
CEA Paris-Saclay, Institut de Recherche sur les lois Fondamentales de l’Univers, 91191 Gif sur Yvette, France; {\it{jesse.palmerio@cea.fr,tatyana.sadibekova@cea.fr}}\\ 
\and 
Université Paris-Saclay, Université Paris Cité, CEA, CNRS, AIM, 91191 Gif-sur-Yvette, France;
\and 
LUX, Observatoire de Paris -- PSL, CNRS, Sorbonne Universit\'e, 92190 Meudon, France \\
\and
UNIDIA, Observatoire de Paris -- PSL, CNRS, 92190 Meudon, France \\
   {\small Received 202x month day; accepted 202x month day}}

\abstract{
The VT--VHF data comprise three types of onboard-processed data results generated from four sequential observational sequences and transmitted to the ground via a Very High Frequency (VHF) downlink. On the ground, these data are processed by three successive pipelines: the pre-processing pipeline, the VT--VHF data processing pipeline (VVPP), and the VT afterglow candidate pipeline (VTAC). These pipelines perform packet decoding, astrometric and photometric calibration, and afterglow candidate identification, respectively. 
This paper describes the architecture and operational implementation of the VT--VHF ground processing system and assesses its end-to-end performance using the first year of SVOM operations. These data enable rapid identification of GRB optical counterparts. Early detections, while the source is still optically bright, facilitate spectroscopic redshift measurements. Dual-band colors provide preliminary redshift constraints and help identify high-redshift candidates, whereas non-detections in both bands may indicate very high redshift, significant extinction, or intrinsically dark bursts. In-orbit operations show that the VT--VHF ground processing system successfully identifies optical afterglow candidates for a significant fraction of ECLAIRs triggers with available VT--VHF data, demonstrating its robustness and readiness.
}

   \authorrunning{C. Wu, J. T. Palmerio \& T. Sadibekova et al.}            
   \titlerunning{SVOM/VT: On-ground processing of VT--VHF data}  

   \maketitle

\section{Introduction}           
\label{sect:intro}
Gamma-ray bursts (GRBs) are among the most energetic cosmic events and provide insight into extreme astrophysical conditions and the evolution of the early Universe. Rapid and accurate localization of GRBs and their afterglows is therefore essential for enabling prompt, multi-wavelength follow-up observations, which are key to constraining their physical nature and evaluating their potential as cosmological probes. 
To address this requirement, 
the SVOM (Space-based multi-band Variable Object Monitor) mission \citep{2016arXiv161006892W} employs a synergy of instruments in space and on the ground. 
Once the onboard telescope for the detection and localization of the GRB prompt emission, ECLAIRs \citep{ECL2025svom} triggers an event exceeding the automatic satellite slewing threshold, the spacecraft repoints within five minutes to bring the X-ray and optical telescope the MXT \citep{MXT2025svom} and VT \citep{VT-Qiu-GenReview2025svom} onto target, starting observations; based on in-orbit data, approximately 58\% of triggers lead to this autonomous sequence. 
Then, following a preliminary onboard processing, the resulting VT data are rapidly downlinked via Very High Frequency (VHF) ground antenna network \citep{cordier2025svom} and processed on the ground to identify optical counterparts. In the following, we refer to such data as VT-VHF data.
This whole procedure is designed to facilitate prompt follow-up observations with other telescopes, particularly optical spectroscopy requiring arcsecond localization.

The generation and content of the VT--VHF data are determined by the predefined onboard VT processing configuration following a GRB trigger.
Four VT observation sequences are scheduled over two post-trigger orbits, starting after platform stabilization following the slew and extending into the subsequent orbit. Each sequence processes 1–6 consecutive high quality images; Seq. 1 and ~2 are acquired consecutively within the first orbit, while Seq.~3 and~4 correspond to the beginning and the end of the second orbit, respectively, with sequences omitted when no valid data are available due to background conditions, Earth occultation, or SAA interruptions.

The onboard VT–VHF pipeline performs basic image calibration, onboard stacking, source extraction, and quality screening to generate three types of data from four VT image sequences (Seq.~1–Seq.~4; \citealt{VTOnboard2025svom}). These include the attitude chart (\texttt{AChart}), which enables onboard pointing reconstruction from a filtered star catalog and that is derived from the first frame of each sequence; the finding chart (\texttt{FChart}), listing point sources extracted from windowed images defined by the GRB localization uncertainties from ECLAIRs or MXT; and the 1-bit subimage (\texttt{SubIm}), obtained by thresholding the original 16-bit windowed images into a binary map that preserves source positional and morphological information while substantially reducing the data volume.
Both the \texttt{FChart} and \texttt{SubIm} are generated from stacked six-frame images to achieve deeper detection limits \citep{Maoqingyun2022SPIE12181E..5SM}.

This paper focuses on the on‑ground processing of the VT–VHF data. The overall workflow and core algorithms for these processing stages are detailed in Section \ref{sect:pipelines}. Section \ref{sect:identification} reviews the practical methodologies and operational experience gained during the commissioning phase for GRB counterpart detection using the VT–VHF data, emphasizing key insights and subsequent refinements to the analysis pipeline. The system's latency, along with its astrometric and photometric performance, is evaluated in Section \ref{sect:preformance}. Finally, Section \ref{sect:conclusion} presents the conclusion, summarizing the scientific significance of the VT–VHF data stream and discussing its potential for enabling rapid, ground‑based identification of optical counterparts in future GRB events.

\section{VT--VHF Ground Processing Pipeline }
\label{sect:pipelines}
The on-ground processing of VT--VHF data involves three primary stages: an initial pre-processing phase that decodes VHF packets (Section~\ref{subsec;pre-proessing}); the VT--VHF data Processing Pipeline (VVPP, Section~\ref{subsec:VVPP}), responsible for astrometric and flux calibrations as well as flagging source attributes; and the VT Afterglow Candidate Pipeline (VTAC, Section~\ref{subsec:VTAC}), designed to identify potential optical counterparts of GRBs. The names of the data products associated with each pipeline, along with the descriptions of their corresponding attributes, are listed in Table~\ref{tab:products}.

The VT--VHF pipelines are part of the large infrastructure at French Science Centre\footnote{https://fsc.svom.org/} (FSC, \citealt{FSC2025svom}). All data processing pipelines are deployed in a containerized environment based on Docker images that adhere to strict quality standards. Services are grouped into stacks and managed via Docker Swarm. Continuous integration workflows ensure that software is tested, validated, and seamlessly deployed across the infrastructure. The orchestration system coordinates the different services and triggers automated stages, from ingesting raw data to pre-processing and generating science products. The environment is deployed across cloud resources in France, at IJCLab\footnote{https://www.ijclab.in2p3.fr/} (The Laboratory of the Physics of the two infinities Irène Joliot-Curie, Orsay) and CC-IN2P3\footnote{https://cc.in2p3.fr/} (CNRS Computing Centre, Lyon), where additional services are maintained with CC-IN2P3 support. This container-based architecture provides the FSC pipelines with reproducibility, scalability, and robust integration into the broader system.

\begin{figure*}
   \centering
   \includegraphics[width=\textwidth]{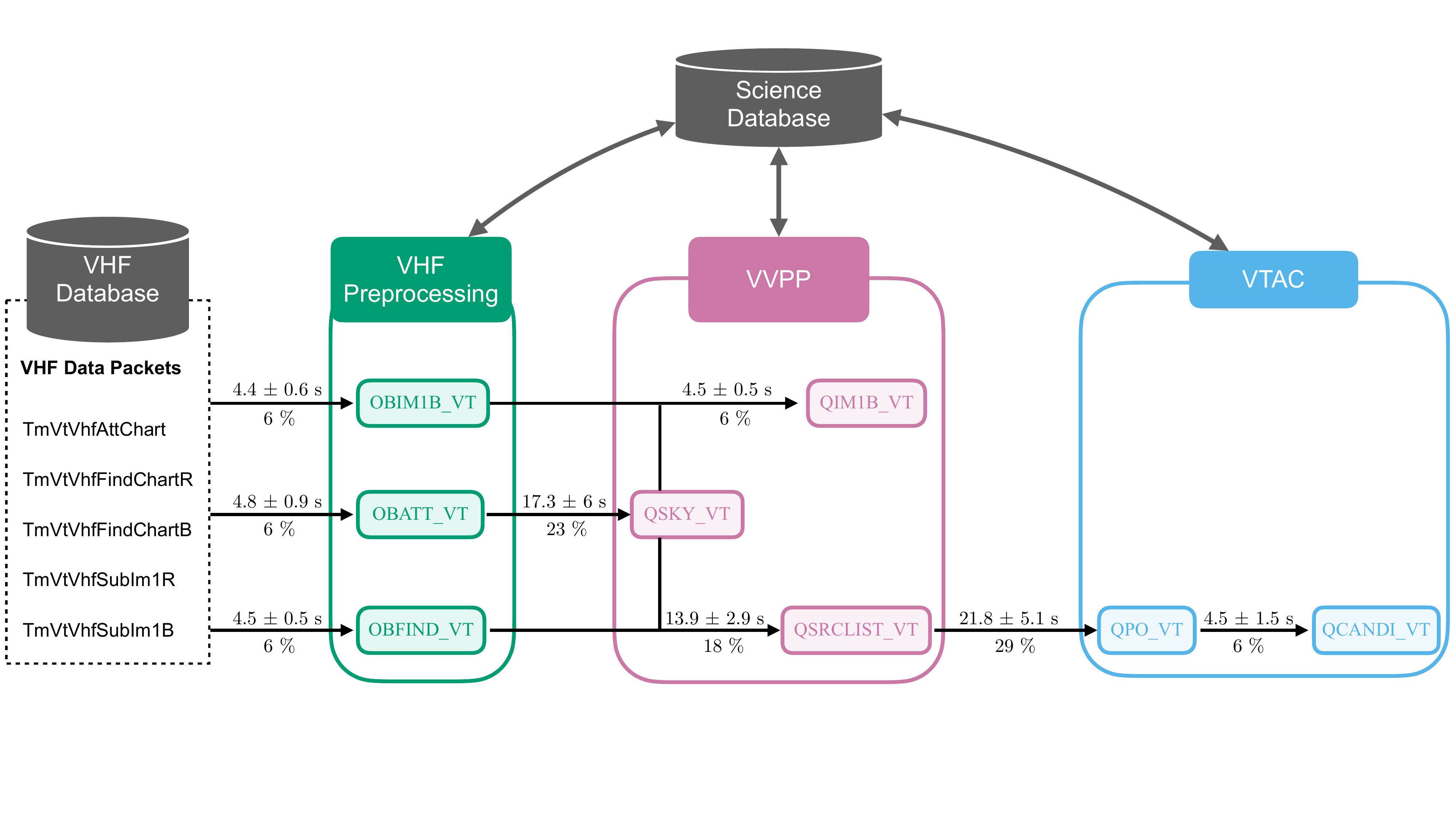}
   \caption{VT--VHF pipeline workflow diagram. All products from each pipeline and their connections are shown. Interactions between products are handled through the database.  For each product, the mean processing time (runtime) $\pm$ standard deviation in seconds is indicated above the branch. The percentage shown below each branch represents the fraction of the total processing time accounted for by that product. These statistics are computed from a sample of 25 confirmed GRBs.}
   \label{Fig-runtime-workflow}
\end{figure*}

Automatic processing is triggered once a generated product relevant to the pipeline and associated with a given event is loaded into the Science Database (SDB). For each event, processing may be triggered multiple times as new VT--VHF data sequences are received. Each pipeline stage in the chain produces its own specific scientific products. The processes and products are described in detail further on in the text. Figure \ref{Fig-runtime-workflow} shows the diagram of the complete processing chain with the complete VT--VHF data flow. In total, eight distinct science products are generated and stored in SDB (listed in Table \ref{tab:products}). 

\begin{table}[t]
\centering
\caption{List of VT--VHF data products}
\label{tab:products}
\small
\begin{tabular}{@{}p{1.8cm} p{3.5cm} p{2cm}@{}}
\hline
Product & Description & Pipeline \\
\hline
OBATT\_VT    & \texttt{AChart} of both channels & pre-processing \\
OBFIND\_VT   & \texttt{FChart} of both channels & pre-processing \\
OBIM1B\_VT   & \texttt{SubIm}  of both channels & pre-processing \\
QSKY\_VT     & Astrometric calibration file & VVPP \\
QSRCLIST\_VT & Calibrated \texttt{FChart} of both channels  & VVPP \\
QIM1B\_VT    & Calibrated \texttt{SubIm} of both channels  & VVPP \\
QPO\_VT      & Position of afterglow candidates  & VTAC \\
QCANDI\_VT   & Properties of afterglow candidates  & VTAC \\
\hline
\end{tabular}
\end{table}

\subsection{Pre-processing Pipeline}
\label{subsec;pre-proessing}
   The pre-processing pipeline is designed to convert raw mission data into standardised products in near-real-time, ready for scientific analysis. The satellite transmits mission data in packets through the VHF ground antenna network  \citep{cordier2025svom}. A dedicated VHF packet management service uploads, decodes, and stores the data at the FSC. Once the data is available at the FSC, the VHF pre-processing pipeline begins downloading the data sequences and running automated processing. The pipeline performs the reconstruction of fragmented data, writes them in standardised output, and ensures a near-real-time data availability. Data transmitted through the VHF channel is the result of the quick onboard analysis.

The pipeline module responsible for managing VT data is named $obloc\_vt$. The module processes three different types of VT content: attitude charts, finding charts, and 1-bit sub-images in the red (VT\_R) and blue (VT\_B) bands, as defined for the VT instrument \cite{VT-Qiu-GenReview2025svom}. Each is assigned a different onboard identification name (APID): \texttt{TmVtVhfAttChart}, \texttt{TmVtVhfFindChart}, \texttt{TmVtVhfSubIm1R}, \texttt{TmVtVhfSubIm1B}. This allows the content to be automatically grouped by type on the ground.  Each type of data is associated with its respective product. The module generates three products for a given alert: \texttt{OBATT\_VT}, \texttt{OBFIND\_VT} and \texttt{OBIM1B\_VT} and continuously updates them with every new sequence. These products are written in the astronomy-specific file format, FITS (Flexible Image Transport System, \citealt{1981A&AS...44..363W}) and sent to SDB for further scientific analysis. 
Since no calibration is applied at this stage, these data have no physical units. This preliminary step does not involve scientific analysis; however, various format and data conversions may be applied to streamline downstream processing. The pre-processing stage processes data from all SVOM instruments, including VT.

\subsection{VVPP: VT--VHF data Processing Pipeline}
\label{subsec:VVPP}
\begin{figure}
   \centering
   \includegraphics[width=0.45\textwidth]{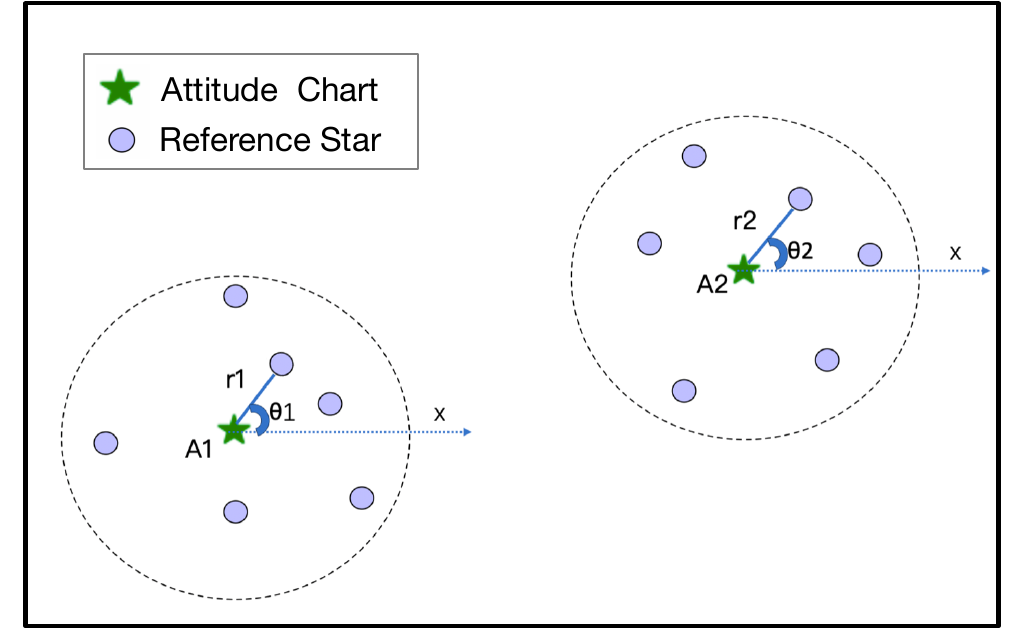}
   \caption{Schematic of the astrometric calibration geometry. For each attitude-chart star (green), reference stars (blue) within a search radius are selected. The vector to each reference star defines a distance $r$ and an angle $\theta$ relative to the $x$-axis, as illustrated for A1 and A2.}
   \label{Fig-pipeline-VVPP-astrometry}
   \end{figure}

     VVPP operates on the pre-processed VT--VHF data products \texttt{OBATT\_VT}, \texttt{OBFIND\_VT}, and \texttt{OBIM1B\_VT}, applying dedicated processing designed for each product type (\texttt{AChart}, \texttt{FChart}, and \texttt{SubIm}) to produce the corresponding outputs \texttt{QSKY\_VT}, \texttt{QSRCLIST\_VT}, and \texttt{QIM1B\_VT}, respectively, as illustrated in Figure~\ref{Fig-runtime-workflow}.
     The VVPP pipeline performs astrometric calibration on the \texttt{AChart} to establish a precise astrometric reference frame. The input catalog is constructed from the 21 brightest stars in the full-frame VT image, with careful filtering to exclude sources exhibiting poor extraction quality. The astrometric calibration is performed by first deriving the pointing information (coordinates $(RA,~Dec)$ and roll angle $\phi$) from the satellite's quaternion parameters and combining it with the known pixel scale to obtain an initial World Coordinate System (WCS), which enables projection of Gaia~EDR3 \citep{gaiaedr32021A&A...649A...1G} reference stars onto the image plane $(x,~y)$. From the \texttt{AChart}, up to 21 stars are selected as matching anchors. For each anchor star, all reference sources within a predefined search radius are retrieved. The relative distance $r$ and position angle $\theta$ between the anchor and each reference star are computed (see the geometric definition in Figure~\ref{Fig-pipeline-VVPP-astrometry}). Candidate pairs are then filtered by requiring consistency in their $(r,\theta)$ vectors, so that true matches share compatible geometric relations while unrelated stars do not. The resulting set of 
reliable matched pairs is supplied to the IRAF \citep{IRAF1986SPIE..627..733T} task \texttt{CCMAP} to derive the transformation between the image coordinates $(x, y)$ and the celestial coordinates $(RA, Dec)$. The resulting astrometric solution is applied to both 
\texttt{FChart} and \texttt{SubIm}.
     
      For \texttt{FChart}, we perform an enhanced astrometric calibration. This process uses its narrower field of view (reducing optical distortion) and higher source density (up to 400 sources) to improve matching. Following a procedure similar to that for \texttt{AChart} and using Gaia EDR3 as reference, we first project the catalog onto the xy-plane using the \texttt{AChart} calibration, then cross-match with \texttt{FChart} detections. The resulting matched list is processed with the IRAF \texttt{CCMAP} task to derive the $(x,~y) \rightarrow (RA,~Dec)$ transformation, finalizing the refined calibration.\footnote{We adopt the IRAF \texttt{CCMAP} task because its astrometric algorithms are well established and robust, and its flexible parameter settings are well suited to our calibration requirements, even though it is part of a legacy software system.}

Following the astrometric calibration, we proceed to the photometric flux calibration.
    Our ground-based flux calibration process utilizes the instrumental magnitudes from onboard processing \citep{VTOnboard2025svom}.
    These onboard magnitudes are derived from raw ADU counts via fixed aperture photometry, prior to exposure time correction. A refined zero point is then applied to standardize the photometric measurements. This zero point  includes  corrections for aperture effects, exposure time, detector gain, and instrumental throughput losses. Its value is primarily determined from the VT \texttt{X-band} data processing \citep{VT-Yao-Calib2025svom}. For rapid processing, a quasi fixed zero point is adopted.

      To assist in identifying GRB optical counterparts, a quality control process is applied to provide empirical assessments of each detection based on its photometric measurements. This empirical method has been validated using the full data sets downloaded through X-band.
      The \texttt{FLAGS}\footnote{It is denoted as ``EFLAG'' (Enhanced Flag) in the final product \texttt{QSRCLIST\_VT}.} parameter is implemented as a bitmask integer encoding essential data quality information for each detected source. Specific flag values indicate distinct conditions: \textbf{0} --- fully reliable measurement; \textbf{1} --- presence of at least one saturated pixel; \textbf{2} --- source affected by a significantly brighter object (e.g., saturation spikes or scattered light); \textbf{4} --- initially blended sources; and \textbf{8} --- suspected cosmic-ray detections. Multiple conditions can occur simultaneously and are combined using bitwise OR, allowing compact and efficient representation of complex data quality scenarios.

     The processing of \texttt{SubIm} involves applying the astrometric solution from the \texttt{QSKY\_VT} product to the corresponding \texttt{QIM1B\_VT} data, writing the calibration results into the \texttt{SubIm} FITS header as a WCS, and updating the header according to SVOM data product standards. This provides timely, though degraded image based information for the identification of GRB optical counterparts.
      
\subsection{VTAC: VT Afterglow Candidate pipeline}
\label{subsec:VTAC}
   The VTAC pipeline serves as the final component in the processing chain for VHF data from VT. Its goal is to analyze the source list (\texttt{QSRCLIST\_VT}) produced by the VVPP pipeline, and optionally leverage positional information from the MXT (\citealt{MXT2025svom}) position product (\texttt{QPO\_MXT}), to identify optical counterpart candidates.
    VTAC generates two data products: the VT Quick Position (\texttt{QPO\_VT}) and the VT Quick Candidate (\texttt{QCANDI\_VT}).
    
    The \texttt{QPO\_VT} product reports the unique sources detected over the course of an alert, together with their positions, magnitudes, and variability indicators. Each source is then classified into one of four tiers S, A, B and C described below, characterizing its likelihood of being a GRB optical counterpart. The detailed values used for each criteria are given in Table~\ref{tab:vtac_candidate_tiers}.

   \begin{table}[t]
    \centering
    \caption{Criteria and thresholds used by VTAC for categorizing candidates in S, A or B tiers. For the catalog criteria, $\Delta\rm m$ is defined as $\rm m_{VT} - m_{cat}$ with the catalog magnitude chosen to be the closest to the VT bands. At the moment, there is not conversion between photometric systems.}
    \label{tab:vtac_candidate_tiers}
    \small
    \begin{tabular}{| c | c c c |}
    \hline
    Tier & Magnitude & Variability & Catalog \\
    \hline
    
    \multirow{2}{*}{S} & in both bands &  & Uncataloged or\\
    & $\rm m < 21$ & $|\Delta\rm m| > 0.1$ \& $|\Delta\rm m| > 5\sigma$ & $\Delta\rm m > 5$ \& $|\Delta\rm m| > 5\sigma$ \\
    \hline
    
    \multirow{2}{*}{A} & & & Uncataloged or \\
    &  $\rm m < 22$ & $|\Delta\rm m| > 3\sigma$ & $\Delta\rm m > 3$ \& $|\Delta\rm m| > 3\sigma$ \\
    \hline
    
    \multirow{2}{*}{B} & & \multirow{2}{*}{$|\Delta\rm m| > 3\sigma$} & \multirow{2}{*}{$\Delta\rm m > 0$ \& $|\Delta\rm m| > 3\sigma$} \\
     & & & \\
    
    \hline
    \end{tabular}
    \end{table}

    \begin{itemize}
        \item \textbf{S tier}: The most confident optical counterpart candidates. Sources in this class meet strict selection criteria, ensuring almost certain identification as a counterpart. They must be detected in both bands, be brighter than magnitude 21, show a significant brightness variation over time, and either not be listed in any catalog or be significantly brighter than the corresponding reference source.
        
        \item \textbf{A tier}: High-probability optical counterpart candidates. The selection criteria are slightly less strict than for the S tier. Sources must be detected in at least one band, show a significant brightness variation over time, and either not be listed in any catalog or be significantly brighter than the corresponding reference source.
        
        \item \textbf{B tier}: Lowest confidence candidates, requiring careful review. Sources must be detected in at least one band, show a significant brightness variation over time, and be listed in at least one catalog.
        
        \item \textbf{C tier}: Sources not considered as candidates. This tier includes all objects that do not meet the criteria for S, A, or B, and are therefore unlikely to be optical counterparts.
        
    \end{itemize}

This classification process is performed by the Identification Algorithm, which evaluates several properties of each detected source. Two main criteria are considered:
    \textit{(i)} \textbf{Photometric behaviour}: The algorithm analyses the measured magnitude and its variations across the observation sequences. Multiple thresholds are applied to assess the significance of both the magnitude measurements and their temporal changes.
 \textit{(ii)} \textbf{Catalog crossmatching}: Each source is compared with known objects from online catalogs. For matched sources, the magnitude difference is computed and evaluated using predefined significance thresholds and a catalog-dependent crossmatch radius. The catalogs used in this process include Gaia, Pan-STARRS, SDSS, and the Legacy Survey.

The \texttt{QCANDI\_VT} product summarizes properties and computes light-curves and temporal indexes for the candidates classified in S, A and B tiers.

\section{Optical Counterpart Validation }
\label{sect:identification}
The Identification Algorithm implemented in VTAC serves as a preliminary selection tool for identifying potential optical counterparts. The results obtained by the algorithm are then reviewed by the VT Instrument Scientist on duty at the time of the alert, who aims to identify the true optical counterpart among the list of candidates.

The validation phase mainly aims at discarding false candidates produced by the pipeline, mostly resulting from instrumental artifacts or observational effects. The most frequent cases identified are as follows: \textit{(i)} blooming\footnote{\url{https://evidentscientific.com/en/microscope-resource/knowledge-hub/digital-imaging/concepts/blooming}} from bright sources can obscure or distort faint objects, with the effect being significantly stronger in VT\_R than in VT\_B;
\textit{(ii)} hot pixels are a source of false candidates and occur more frequently in VT\_B than in VT\_R. This higher rate in VT\_B is due to the comparative lack of the radiation hardening features present in VT\_R, namely its lower operating temperature and deep-depletion CCD design \citep{VT-Qiu-GenReview2025svom}.
\textit{(iii)} cosmic rays may remain despite each sequence being extracted from a stack of 3 individual frames to allow for easier removal, such cases are usually identified using the \texttt{EFLAG} parameter.

After these cleaning steps, two additional checks are performed: \textit{(i)} remaining sources are compared with known Solar System objects and minor planets to ensure that no transient is due to a passing object;
\textit{(ii)} candidate positions are cross-checked with observations from other instruments (such as X-ray observations from SVOM/MXT, \textit{Swift}-XRT or \textit{EP}/FXT). When available, multi-wavelength detections should be spatially consistent with the candidate’s position.

Our findings from nearly a year of operations indicate that this validation process is reliable for sources brighter than 20~mag, whereas fainter sources require greater caution and are best validated through combined multi-sequence analysis.
      
\section{System Performance Analysis }
\label{sect:preformance}

   From the launch of SVOM to 2025 September 3 (the last event: sb25090304/GRB~250903A, \citealt{2025GCN.41677....1M}), covering the entire commissioning phase, ECLAIRs, including joint triggers with the GRM, detected and confirmed 55 GRBs. Among these, VT performed automatic slewing and follow-up observations for 32 events, i.e., those that surpassed the onboard slewing threshold. Of the 32 slews, 25 generated VT–VHF data products\footnote{This count includes only the \texttt{AChart} products, which provide photometric and astrometric information of the objects.}
    Within this subset of 25 events with usable data, optical counterparts were identified in 14 cases. Of these 14 counterparts, 9 were promptly reported to the GCN. The remaining 5 cases did not lead to a confident identification of an optical counterpart, due to issues such as low SNR or localization uncertainties from the high-energy instrument.
   The loss or partial loss of VT–VHF data can be attributed to multiple factors, including stray light and other environmental effects that degraded data quality and led to onboard processing failures, as well as limitations in the onboard logic and parameter settings. With progressive system optimization during commissioning, the failure rate is expected to decrease further in future operations. A statistical summary of these samples is presented in Table \ref{Tab-grb_sample_distribution}.

\begin{table*}[htbp]
\centering
\caption{Distribution of GRB Observations: VT--VHF Data Sample ($n = 55$)}
\begin{tabular}{lrrrr}
\hline
Sample Category & $n$ & \% of Total & \% of Autoslewing & \% of Parent \\
\hline
Total Sample & 55 & 100.0 & -- & -- \\
\hline
Autoslewing GRBs & 32 & 58.2 & 100.0 & -- \\
\quad -- VHF data available & 25 & 45.5 & 78.1 & 78.1 \\
\quad \quad -- Optical counterpart identified & 14 & 25.5 & 43.8 & 56.0 \\
\quad \quad \quad -- Reported to GCN & 9 & 16.4 & 28.1 & 64.3 \\
Non-Autoslewing GRBs & 23 & 41.8 & -- & -- \\
\hline
\end{tabular}
\label{Tab-grb_sample_distribution}
\end{table*}

\subsection{Timing Performance Analysis}
\label{subsec:runtime-performances}
To evaluate the capability of the ground-based processing pipeline in delivering rapid responses for GRB optical counterpart detection, we performed a runtime analysis of its three main stages : VHF pre-processing, VVPP, and VTAC. For this analysis, we specifically focused on the subset of confirmed GRBs with VT--VHF data products generated. This subset includes 25 events and is specifically chosen to best  highlight  the effectiveness of the pipeline for its primary scientific objective.

Figure~\ref{Fig-runtime-workflow} shows the mean runtime and variability for each product. Most of them have stable runtimes $\sim$5~s with low variability, indicating efficient and consistent performance. In contrast, some products, such as \texttt{QSRCLIST\_VT} and \texttt{QPO\_VT}, are more time-consuming and variable since their runtimes depend strongly on the number of received observation sequences. Despite that, runtimes are still below 30~s, which is highly acceptable.

Table~\ref{Tab-runtime-contribution} summarizes the mean runtime and relative contribution of each processing stage within the pipeline. On average, the complete VT--VHF on-ground processing requires $\sim$76~s, with VVPP representing the largest share (47.1\%). These runtimes remain relatively short, allowing scientific analysis in about a minute after data acquisition. Compared to other system delays, such as the satellite VHF downlink, ground processing does not appear as a limiting factor for overall response time.

\begin{table}[h]
    \centering
    \caption[]{Mean Time Contribution of each pipeline to the SVOM/VT--VHF Data Ground Processing. Runtimes are calculated on a total of 25 confirmed GRBs.}
    \label{Tab-runtime-contribution}
    \setlength{\tabcolsep}{1pt}
    \begin{tabular}{
        >{\centering\arraybackslash}m{1.8cm} |
        >{\centering\arraybackslash}m{1.5cm}
        >{\centering\arraybackslash}m{1.5cm}
        >{\centering\arraybackslash}m{1.5cm}
        >{\centering\arraybackslash}m{1.5cm}
    }
        \hline\noalign{\smallskip}
        & pre-processing & VVPP & VTAC & \textbf{Total} \\
        \hline\noalign{\smallskip}
        Mean $\pm$ std runtime (s) & 
        13.7 $\pm$ 1.2 & 
        35.7 $\pm$ 6.7 & 
        26.4 $\pm$ 5.3 & 
        \textbf{75.8 $\pm$ 8.6} \\
        \noalign{\smallskip}\hline
    \end{tabular}
\end{table}

\begin{figure*}
   \centering
   \includegraphics[width=1\textwidth]{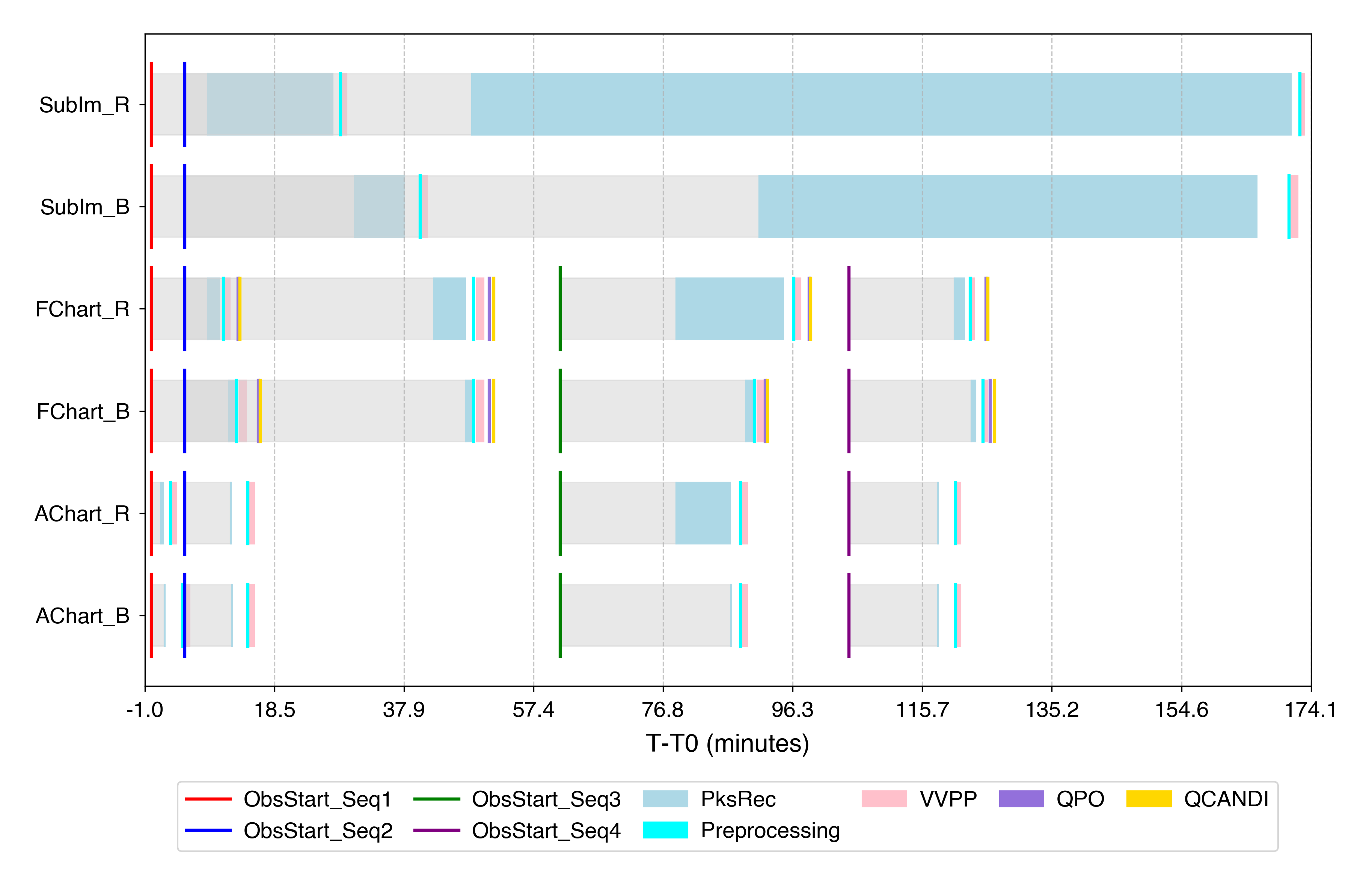}
\caption{
A typical example (sb25080601/GRB~250806A \citet{2025GCN.41243....1X}) of the VT--VHF data processing timeline. 
The colored bars mark the start times of the VT observations for each of the four sequences, while the gray bars indicate the durations of exposure, onboard processing, and data packet buffering and transmission following each observation onset. 
The buffering time varies with data priority, whereas transmission time depends mainly on network traffic and may overlap with data from other onboard instruments. 
This example illustrates the overall latency and variability of product generation for different VT data types (\texttt{AChart}, \texttt{FChart}, and \texttt{SubIm}).
}
\label{Fig-timeline-visual}
\end{figure*}

For scientific users, the key concern is the end-to-end latency, namely the total time from the $T_{0}$ trigger through spacecraft slew and pointing, onboard processing, data downlink, and on-ground processing to the delivery of processed data products to ground users. 
To illustrate this, Figure~\ref{Fig-timeline-visual} presents a representative example of the VT–VHF processing timeline (sb25080601/GRB~250806A; \citealt{2025GCN.41243....1X}), showing the temporal evolution of the three VT–VHF data products across four observation sequences, from the start of exposure to the completion of the final outputs. In this GRB case, VT began observations at $T_{0}+3.77$~min, with each sequence containing six 50~s frames (a total exposure of 300~s)\footnote{The \texttt{AChart} product does not stack frames; it uses only the first 50~s exposure of each sequence.}.
Gray bars in Figure~\ref{Fig-timeline-visual} indicate the durations of exposure, onboard processing, and data packet buffering and transmission; the buffering time varies depending on data priority. As shown in the figure, from the start of observation to the completion of final products, the total processing time ranges from 3.74 to 28.03~min for \texttt{AChart}, 13.35 to 46.45~min for \texttt{FChart}, and 29.3 to 168.09~min for \texttt{SubIm}. These variations primarily arise from sequence-dependent packet weighting, which determines transmission priority and consequently affects buffering time. Another significant factor is the data packet transmission time, which depends on the level of network traffic. 
This representative event demonstrates that users can typically access the \texttt{FChart} and \texttt{SubIm} products through the VT--VHF pipeline within about 30~min after $T_{0}$. These products enable a preliminary identification of the GRB optical counterpart and its positional association in the 1-bit subimages.

  Within this end-to-end timeline, a practical question is how soon after $T_{0}$ VT can begin its first valid observation. We term this interval the effective slew delay, including both the spacecraft slew to the target and the subsequent validation of image quality against onboard processing thresholds. Of the 25 events with VT--VHF data, 20 produced both VT\_B and VT\_R products in the Seq.~1 series. Their effective slew delays  have a minimum of 243 s and a median of 325 s, with all but one outlier (extended to 898 s due to high background) below 460 s. For the remaining five events, VT was near or outside its observing window during the trigger orbit\footnote{Although ECLAIRs issued the trigger within its own observable window, VT was either at the end of its observing window or had already entered an unobservable region, with straylight issue or preventing same orbit pointing.}. Consequently, the first available data in the following orbit appear as Seq.~3, and no Seq.~1 data were obtained for these events.

 \subsection{Astrometric and Photometric Performance Analysis}
  \label{subsect:sci-analysis}

   \begin{figure}
   \centering
   \includegraphics[width=0.4\textwidth]{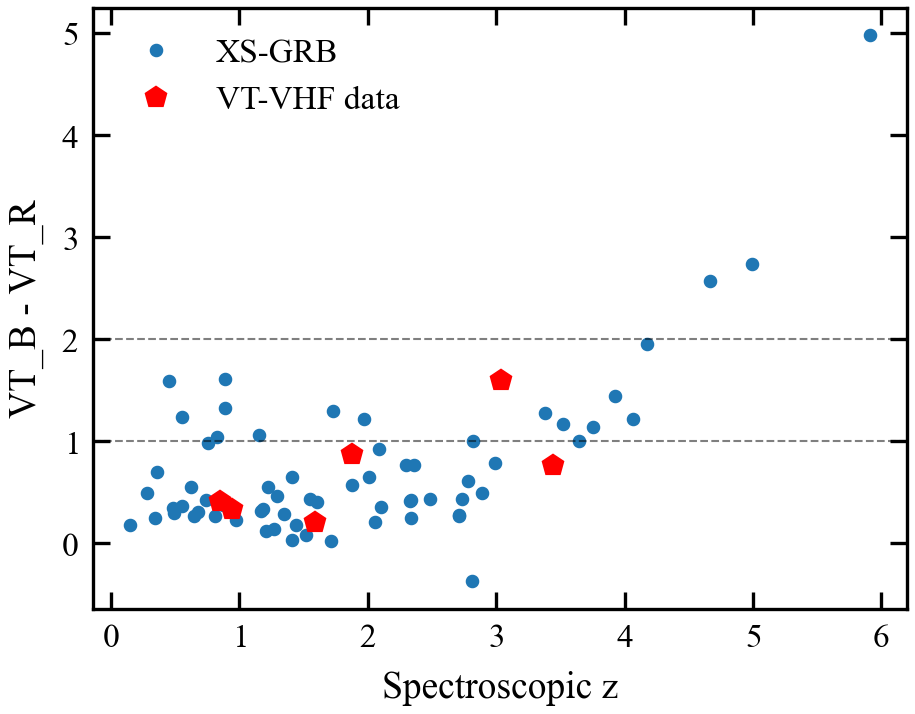}
\caption{Color--redshift relationship of GRB afterglows. The blue circles represent the XS-GRB sample\citep{2019A&A...623A..92S}, showing the $(\mathrm{VT\_B} - \mathrm{VT\_R})$ color as a function of spectroscopic redshift. The magnitudes are derived from simulated spectra convolved with the VT filter response curves. The red pentagons represent the corresponding measurements from real VT--VHF observations. The horizontal dashed lines indicate the adopted color cuts at \( \mathrm{VT\_B - VT\_R} = 1 \) and \( = 2 \), which are used as empirical criteria to assist the identification of potential high-redshift GRB candidates based on color information.}
   \label{Fig-color-shift}
\end{figure}

The accuracy of astrometric and flux calibration represents a key metric for assessing the reliability of GRB counterpart identification from VT--VHF data. For the 25 events with available VT--VHF data (\texttt{AChart} products), the astrometric solution calibrated against the Gaia EDR3 catalog yields a median 1$\sigma$ fitting residual of $0.024\arcsec$, with 90\% of solutions better than $0.056\arcsec$. No significant difference is found between the two bands, indicating that the astrometric calibration is consistent across VT\_B and VT\_R.

Cross-validation with Gaia EDR3 shows that the calibrated \texttt{FChart} catalog has a positional accuracy of $0.20\arcsec$ overall for both channels (90\% C.L.). For sources at 20th magnitude\footnote{This magnitude marks the empirical brightness limit for credible counterpart validation with VT–VHF data, which is brighter than the VT’s detection limit.}, the positional accuracy is $0.33\arcsec$. For these faint VT sources, the positional error is dominated by source detection (SNR-limited), not by astrometric calibration, making this value a typical upper limit.

Our flux calibration analysis, conducted through comparison with ground-based VT \texttt{X-band} data processing \citep{VTX-BAND2025svom}, reveals systematic magnitude offsets of $-0.37 \pm 0.18$\,mag in VT\_B and $-0.12 \pm 0.13$\,mag in VT\_R over more than one year of in-orbit observations. These measured offsets demonstrate long-term instrumental variations that necessitate regular zeropoint updates based on continuous monitoring. It should be noted, however, that these systematic biases and dispersions do not affect the determination of GRB luminosity variations. This is because the relative photometry derived for individual GRBs from VT–VHF data remains stable on the short timescales characteristic of burst observations.
 
One notable advantage of the VT--VHF data is its ability to provide rapid color measurements shortly after the GRB trigger. To explore the relation between color and redshift, all six GRB afterglows in the VT--VHF sample with both spectroscopic redshift and color information are shown in the color--redshift diagram (Figure~\ref{Fig-color-shift}), following the method of \citet{2020RAA....20..124W}.
As illustrated in Figure~\ref{Fig-color-shift}, the colors derived from VT--VHF observations appear broadly consistent with the empirical trend reported by \citet{2020RAA....20..124W}, in which the color becomes redder at \( z \gtrsim 3 \). This trend is mainly attributed to the redshifted Lyman break caused by absorption from intervening neutral hydrogen along the line of sight. However, dust extinction in the GRB host galaxy may also contribute to redder observed colors, particularly for events at moderate redshifts.  
We note that the present analysis is based on a limited sample of only six GRBs with both spectroscopic redshift and color measurements. As such, the observed consistency should be regarded as indicative rather than statistically conclusive. Despite this potential contamination, the agreement between the simulated relation and the observational data suggests that color information can provide a useful, though coarse, indication of redshift. In this context, relatively red colors (e.g., \( \mathrm{VT\_B - VT\_R} \gtrsim 2 \)) are more often associated with high-redshift GRBs ($z \gtrsim 4$), though dusty lower-redshift events cannot be excluded, while bluer colors (e.g., \( \mathrm{VT\_B - VT\_R} \lesssim 1 \)) are more commonly found at lower redshifts ($z \lesssim 4$). Future expansion of the VT--VHF sample will be essential to verify and refine this color--redshift relation with improved statistical significance.

\section{Conclusions and perspectives}
\label{sect:conclusion}
The VT--VHF on-ground processing framework, operating in conjunction with onboard processing and the VHF communication network, demonstrates the capability of VT for rapid GRB follow-up and optical counterpart identification. Based on the first year of in-orbit operations, the processing chain typically delivers the Fchart and SubIm products from the initial observation sequence within $\sim 30$ minutes post-trigger. The SubIm products are generated as soon as either band is processed, enabling a preliminary identification of the GRB optical counterpart and its positional association in the 1-bit subimages. The resulting localization accuracy is better than 0.5~arcsec. The median processing time of the VT--VHF on-ground pipeline is $\sim 76$~s, indicating efficient data handling. For VT--VHF data sets with detected optical afterglows, the current identification success rate is 64.3\%, providing a solid baseline for mission performance. Unsuccessful cases are primarily associated with faint afterglows or detections limited to a small number of sequences or bands, leading to insufficient confidence for reliable identification.

Further improvements are expected through continued refinement of the VT--VHF on-ground processing algorithms to enhance identification efficiency and robustness against contaminants such as hot pixels, cosmic rays, and nearby bright stars. In parallel, ongoing optimization of operational strategies and overall system performance is anticipated to improve VHF data availability and completeness, thereby strengthening the foundation for higher automatic follow up and afterglow identification success rates in future operations.

\begin{acknowledgements}
The Space-based multi-band astronomical Variable Objects Monitor (SVOM) is a joint Chinese-French mission led by the Chinese National Space Administration (CNSA), the French Space Agency (CNES), and the Chinese Academy of Sciences (CAS). We gratefully acknowledge the unwavering support of NSSC, IAMCAS, XIOPM, NAOC, IHEP, CNES, CEA, and CNRS. We also gratefully acknowledge support from the SVOM FSC engineer team and CNRS/IN2P3 Computing Center (Lyon - France) for providing computing and data-processing resources needed for this work. This work is supported by the Strategic Priority Research Program of the Chinese Academy of Sciences (Grant No.XDB0550401), and by the National Natural Science Foundation of China (grant Nos. 12494571, 12494570 12494573, and 12133003).
The authors are thankful for support from the National Key R\&D Program of China (grant Nos. 2024YFA161170* and 2024YFA1611700).
\end{acknowledgements}

\appendix                  

\bibliography{ms2026-0022}{}

\label{lastpage}
\clearpage

\bibliographystyle{raa} 

\end{document}